\documentclass{aa}  

\usepackage{txfonts}
\usepackage{tabularx}
\usepackage{subfigure}  
\usepackage{graphicx,amsmath}
\usepackage{amssymb}
\usepackage{color}
\usepackage{natbib}      
\usepackage{url}
\usepackage{multirow}
\usepackage[para,online,flushleft]{threeparttable}   
\usepackage{soul,mathabx}
\bibpunct{(}{)}{;}{a}{}{,} 
\usepackage{lscape}
\usepackage{adjustbox,textcomp}
\usepackage{dblfloatfix}
\usepackage{dsfont}

\usepackage{hyperref}  
\hypersetup{
    colorlinks=true,
    linkcolor=blue,
    filecolor=blue,      
    urlcolor=blue,
    citecolor=blue,
    }
\usepackage{braket}     
\usepackage{upgreek}    
\usepackage{array,multirow,booktabs}

\newcommand{\J}{\texttt{JADE}}
\usepackage{xspace}

\begin{document} 

\title{Exploring the parameter space of hierarchical triple black hole systems}

\author{M.~Attia
   \and Y.~Sibony}

\institute{Observatoire Astronomique de l'Universit\'e de Gen\`eve, Chemin Pegasi 51b, CH-1290 Versoix, Switzerland\\
           \email{mara.attia@unige.ch}
          }

\authorrunning{M.~Attia \& Y.~Sibony}
\titlerunning{Hierarchical Triple Black Hole Parameter Space}

\date{Received 15 April 2025 / Accepted 17 June 2025}

\abstract
{
We present a comprehensive exploration of hierarchical triple black hole (BH) systems to address the long-standing ``initial separation'' problem in gravitational wave (GW) astrophysics. This problem arises because isolated BH binaries must have extremely small initial separations to merge within a Hubble time via GW emission alone, separations at which their stellar progenitors would have merged prematurely. Using a modified version of the \J{} secular code that incorporates GW energy loss, we systematically investigate a seven-dimensional parameter space consisting of the masses of the three BHs ($5 - 100\,M_\odot$ for the inner binary components, $1 - 200\,M_\odot$ for the tertiary), inner and outer semimajor axes ($1 - 200\,\mathrm{AU}$ and $100 - 10,000\,\mathrm{AU}$, respectively), outer orbit eccentricity ($0 - 0.9$), and mutual inclination between orbits ($40^\circ - 80^\circ$). We employed an innovative adaptive Markov chain Monte Carlo approach that preferentially samples the transition boundary between merging and nonmerging configurations, allowing us to efficiently map the merger probability landscape with nearly 15 million distinct simulations. Our results reveal that the parameter space regions most conducive to mergers correspond to systems with asymmetric inner binary masses, large inner separations where the von Zeipel--Lidov--Kozai (ZLK) mechanism can operate effectively without being suppressed by relativistic precession, small outer separations providing stronger perturbations, and large outer eccentricities that bring the tertiary closer at pericenter. Merger probability is also globally positively correlated with mutual inclination, with some irregular features departing from monotonicity. For nonmerging systems, we developed a classification scheme based on the presence of GW emission and ZLK oscillations, identifying distinct regions in parameter space for each category and providing a better understanding of the hierarchical triple channel. Additionally, we trained a neural network that predicts merger outcomes with an area under the receiver operating characteristic (ROC)  curve score of 99\% and an overall accuracy of 95\%, increasing to 99.7\% accuracy for the $\sim$\,80\% of predictions made with high confidence. This model enables rapid population synthesis studies without requiring computationally expensive dynamical simulations. We validated our secular approach through comparison with direct $N$-body integrations for select systems, finding good qualitative agreement in merger outcomes for 87\% of test cases, confirming that our methodology effectively captures the essential dynamics governing triple BH evolution, while enabling exploration at an unprecedented scale. Our results provide crucial insights into which configurations of hierarchical triples can resolve the initial separation problem and can serve as viable progenitors for the growing catalog of GW detections.
}

\keywords{black hole physics -- gravitational waves -- stars: kinematics and dynamics -- methods: numerical}

\maketitle

\section{Introduction}
\label{sec:intro}

The detection of gravitational waves (GWs) has not only confirmed their existence, as predicted by general relativity \citep{Einstein1916,Einstein1918}, but has also revolutionized our understanding of compact object dynamics and opened an entirely new observational window to the Universe. With over 90 confirmed detections to date \citep{Abbott2023}, GW astronomy has rapidly evolved from a theoretical construct to a data-rich observational science, profoundly impacting multiple areas of astrophysics from stellar evolution to cosmology \citep[e.g.,][]{Miller2016,Mapelli2018,Mapelli2021a,Mandel2022a,Mandel2022b}. However, these remarkable observations have brought into sharp focus a fundamental challenge in our understanding of compact object binary formation and evolution, commonly referred to as the ``initial separation'' problem or ``final AU'' problem \citep{Mandel2022a}. For two black holes (BHs) to merge within the age of the Universe through GW emission alone, they must begin at an extraordinarily small separation---typically less than $50\,R_\odot$ for $30\,M_\odot$ BHs. This requirement creates a puzzling apparent contradiction: the massive stars that are the progenitors of these BHs have radii that are much larger than this critical separation during significant portions of their evolution. If such massive stars were placed at separations small enough to enable their eventual BH remnants to merge within a Hubble time, the stars themselves would merge long before they could evolve into BHs.

This contradiction has spurred intense theoretical investigation into various formation pathways that might resolve this tension. The proposed mechanisms fall broadly into two categories: stellar evolution solutions and dynamical solutions. Stellar evolution approaches include mass transfer channels. In stable mass transfer \citep{Ionayoshi2017,GallegosGarcia2021,vanSon2022,Briel2023},  one star fills its Roche lobe and undergoes controlled mass exchange with its companion, potentially leading to orbital tightening. In common-envelope evolution \citep{Paczynski1976,Livio1988,Ivanova2013,Postnov2014},  dynamically unstable mass transfer triggers the engulfment of the mass-gaining star within the donor's envelope, where the orbital energy is used to eject the common envelope and dramatically shrink the binary separation. Another stellar evolution pathway is chemically homogeneous evolution \citep{Marchant2016,Mandel2016,deMink2016,duBuisson2020,Riley2021}, where strong internal mixing prevents stellar expansion, allowing initially close binaries to remain compact throughout their evolution; however, this pathway has become increasingly disfavored due to the lack of high-spin detections in the current GW catalog. Dynamical solutions, on the other hand, focus on mechanisms that can bring already-formed BHs closer together, primarily through interactions in dense stellar environments such as globular clusters \citep{Sigurdsson1993,Kulkarni1993,PortegiesZwart2000,Banerjee2010,Morscher2015,Rodriguez2016,Askar2017,Fragione2018,Mapelli2021b} or through secular interactions in hierarchical multiple systems \citep{Antonini2016,Silsbee2017,Hoang2018,Martinez2020,Dittmann2024}. The hierarchical triple scenario, which is the focus of this paper, leverages the fact that a significant fraction of massive stars form in triple and higher-order multiple systems \citep{Duchene2013,Tokovinin2021}. In such configurations, the von Zeipel--Lidov--Kozai (ZLK) mechanism \citep{vonZeipel1910,Lidov1962,Kozai1962} can induce oscillations in the inner binary's eccentricity. During periods of high eccentricity, the pericenter distance becomes extremely small, dramatically increasing GW emission and potentially leading to a merger within the age of the Universe. An important manifestation of this phenomenon is shown for instance in the recent work of \citet{Stegmann2025}, suggesting that the present-day rate of BH--neutron star mergers may be largely driven by ZLK resonance.

While numerous aspects of the hierarchical triple pathway have been studied in recent years \citep{Liu2020,Liu2021,Su2021,Britt2021,Mannerkoski2021,Trani2022,Chandramouli2022,Gondan2023,Lepp2023,Bartos2023,Liu2024,Generozov2024,Hao2024}, the literature lacks a comprehensive exploration of the full parameter space relevant to BH mergers. Previous works have typically focused on specific subsets of the parameter space, particular mass ranges, or limited samples of systems, making it difficult to develop a holistic understanding of which initial configurations are most likely to produce mergers within the age of the Universe. Our primary aim in this work is to conduct a systematic, large-scale investigation of the parameter space of hierarchical triple BH systems, with a specific focus on identifying the boundary between initial configurations that lead to mergers within 14 Gyr and those that do not. By mapping this boundary in detail, we can provide valuable insights into the viability of the hierarchical triple channel as a solution to the initial separation problem and guide future population synthesis studies seeking to predict the rates and properties of GW sources.

To make this extensive parameter space exploration computationally feasible, we employed a secular approximation that averages the equations of motion over both orbital periods, drastically reducing the computational cost compared to direct $N$-body integration. We validated this approach by comparing a subset of our results with fully self-consistent $N$-body simulations. Furthermore, we developed an innovative Markov chain Monte Carlo (MCMC) sampling strategy that preferentially explores the transition region between merging and nonmerging configurations, allowing us to efficiently map the boundary in our high-dimensional parameter space. As an additional contribution, we trained a neural network on the results of our extensive simulations to create a predictive model that can rapidly classify the merger outcome of any hierarchical triple BH system within our parameter space. The aim of this model is to achieve a trustworthy accuracy, while requiring only milliseconds of computation time per system, enabling future population synthesis studies to efficiently investigate the hierarchical triple channel without the computational burden of full dynamical simulations.

This paper is organized as follows: Section~\ref{sec:context} provides a contextual description of the initial separation problem and the ZLK mechanism. Section~\ref{sec:methods} outlines our methodology, including the secular approximation, MCMC sampling technique, and neural network implementation. Section~\ref{sec:results} presents our findings. We compare secular and $N$-body simulations, characterize the parameter space of merging and nonmerging systems, and evaluate the performance of the neural network predictor. Finally, in Sect.~\ref{sec:conclusion} we discuss the limitations of our approach, summarize our conclusions, and explore the implications for future research.

\section{Context}
\label{sec:context}

The merger timescale $\tau_\mathrm{GW}$ of a binary system due to GW emission can be derived from general relativity. For a circular binary with component masses $m$ and $M$, the merger timescale is

\begin{equation}
\label{eq:tgw}
    \tau_\mathrm{GW} = \frac{5}{256} \frac{c^5}{\mathcal{G}^3} \frac{a^4}{(m M)(m + M)},
\end{equation}

\noindent where $a$ is the binary separation, $\mathcal{G}$ the gravitational constant, and $c$ the speed of light. For two $10\,M_\odot$ BHs, the maximum initial separation for which $\tau_\mathrm{GW} \leq 14\,\mathrm{Gyr}$ (approximately the age of the Universe) is about $12\,R_\odot$ (Eq.~(\ref{eq:tgw})). This creates an apparent contradiction: if the progenitor stars were located this close to each other, they would merge before becoming BHs, due to their much larger radii during evolution. Conversely, if located at larger separations to avoid stellar mergers, the resulting BH binary would not merge within the age of the Universe through GW emission alone.

The ZLK resonance provides a potential solution to this problem. In a hierarchical triple configuration, where an inner binary is orbited by a distant third body, this mechanism can induce large-amplitude oscillations in the inner binary's eccentricity, while conserving angular momentum. These oscillations can temporarily drive the inner binary's eccentricity to extremely high values, dramatically reducing the pericenter distance during certain phases of the cycle. Since GW emission is strongly enhanced at small separations, these periodic excursions to high eccentricity can accelerate the orbital decay by orders of magnitude, potentially enabling mergers within cosmological timescales even for systems that begin with large separations. The effectiveness of the hierarchical triple pathway depends on a complex interplay of multiple factors. The ZLK mechanism requires a sufficiently large mutual inclination $i_{\rm mut}$ between the inner and outer orbits, typically between $39.2^\circ$ and $140.8^\circ$ for circular outer orbits \citep{Kinoshita1999}, though this range can expand for eccentric outer orbits. The strength of the perturbation, and hence the maximum eccentricity attainable by the inner binary, depends on the mass ratios, orbital separations, and eccentricities \citep[and references therein]{Naoz2016}. This phenomenon must also be tempered by the influence of relativistic precession of the inner orbit, which acts as a competing mechanism that can quench the ZLK oscillations if sufficiently strong \citep[e.g.,][]{Liu2015}.

\section{Methods}
\label{sec:methods}

\subsection{Secular evolution with the \J{} code}
\label{subsec:jade}

\begin{figure}
    \centering
    \includegraphics[width=\linewidth]{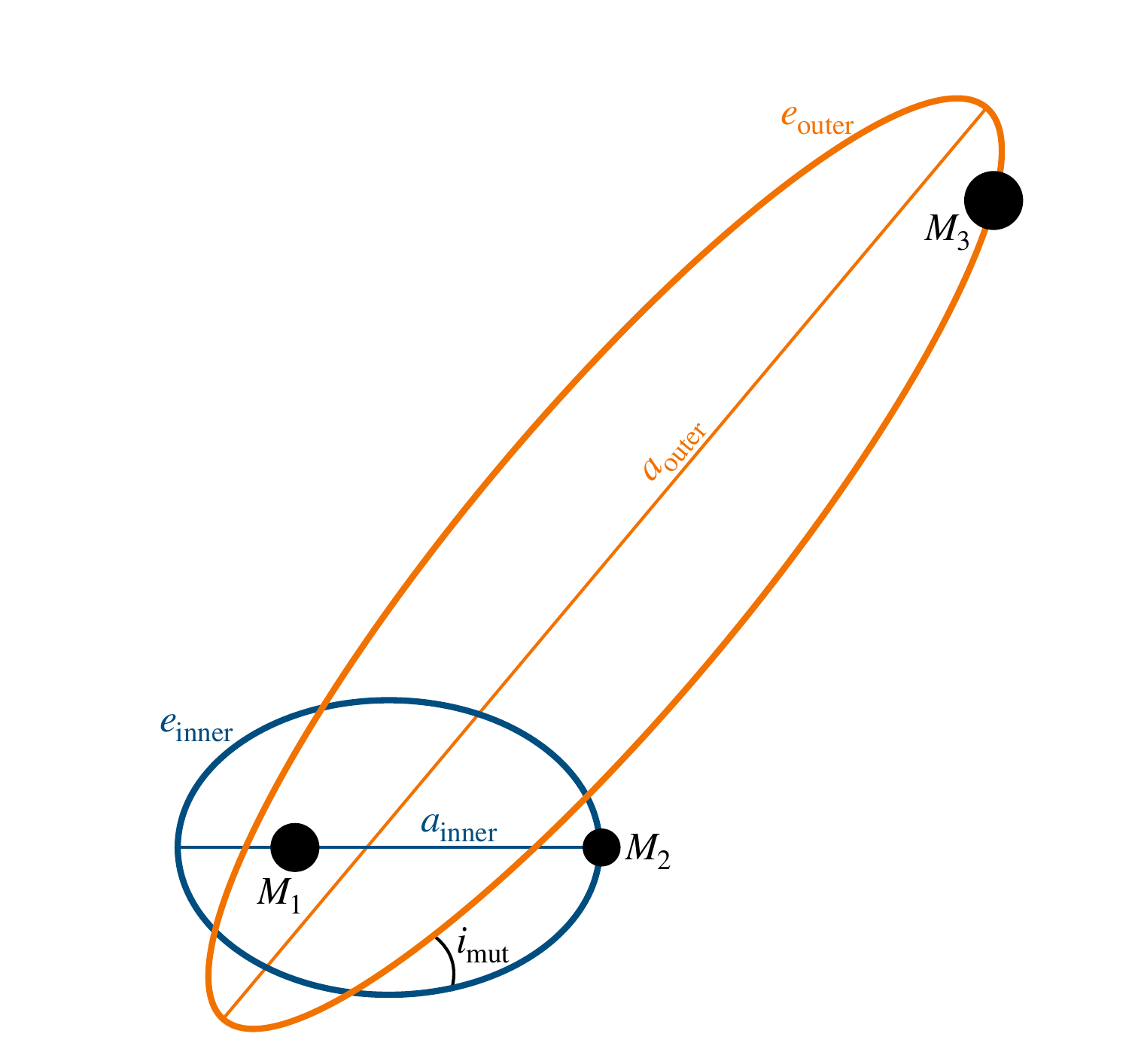}
    \caption{Generic orbital configuration of a hierarchical triple system, as investigated in this study.}
    \label{fig:hierarchical_triple}
\end{figure}

We use a modified version of the \J{}\footnote{\url{https://github.com/JADE-Exoplanets/JADE}} code \citep{Attia2021}, which was originally developed to simulate the secular evolution of hierarchical triple systems in the context of planetary dynamics. The code treats the three-body problem using a secular approximation, where the differential equations of motion are averaged over both orbital motions. This   allows   much longer timesteps and faster computations compared to direct $N$-body integration.

The Hamiltonian of a hierarchical triple system can be expressed as \citep{Harrington1968}

\begin{align}
\begin{split}
\label{eq:ham}
    \mathcal{H} &= \frac{\mathcal{G} M_1 M_2}{2 a_\mathrm{inner}} + \frac{\mathcal{G} M_3 \left(M_1 + M_2\right)}{2 a_\mathrm{outer}}\\
    + \frac{\mathcal{G}}{a_\mathrm{outer}} &\sum_{n=2}^{\infty}{\left(\frac{a_\mathrm{inner}}{a_\mathrm{outer}}\right)^n M_n 
    \left(\frac{\lvert\boldsymbol{r_\mathrm{inner}}\rvert}{a_\mathrm{inner}}\right)^n 
    \left(\frac{a_\mathrm{outer}}{\lvert\boldsymbol{r_\mathrm{outer}}\rvert}\right)^{n + 1} P_n(\cos{\Phi})},
\end{split}
\end{align}

\noindent where $M_1$ and $M_2$ are the masses of the inner binary components, $M_3$ is the mass of the outer companion, $a_\mathrm{inner}$ and $a_\mathrm{outer}$ are the semimajor axes of the inner and outer orbits, $\boldsymbol{r_\mathrm{inner}}$ and $\boldsymbol{r_\mathrm{outer}}$ are the Jacobi coordinates \citep[e.g.,][]{Murray1999}, $\Phi$ is the angle between these vectors, $P_n$ are the Legendre polynomials, and

\begin{equation}
    M_n = M_1 M_2 M_3 \frac{M_1^{n - 1} - (-M_2)^{n-1}}{(M_1 + M_2)^n}.
\end{equation}

\noindent The orbital configuration is depicted in Fig.~\ref{fig:hierarchical_triple}. Following \citet{Beust2012}, \J{} truncates the infinite series of Eq.~(\ref{eq:ham}) at the fourth (hexadecapolar) order, which provides a good balance between accuracy and computational efficiency for hierarchical systems. To the best of our knowledge, this is the highest order that has been investigated for such secular analyses.

In addition to gravitational interactions (Eq.~(\ref{eq:ham})), \J{} also accounts for relativistic precession of the inner orbit, which acts as a quenching mechanism for the ZLK resonance \citep[e.g.,][]{Liu2015}. Furthermore, since the original \J{} code was developed for planetary systems, we modified it to include the effects of GW emission on the inner binary. The loss of orbital angular momentum and eccentricity damping due to GWs are given by \citep{Peters1964}

\begin{equation}
\label{eq:dLdt}
    \left.\frac{\mathrm{d}L_\mathrm{inner}}{\mathrm{d}t}\right|_\mathrm{GW} = -\frac{32}{5}\frac{\mathcal{G}^{7/2}}{c^5}
    \frac{(M_1 M_2)^2 \sqrt{M_1 + M_2}}{a_\mathrm{inner}^{7/2}}\frac{1 + \frac{7}{8}e_\mathrm{inner}^2}{\left(1 - e_\mathrm{inner}^2\right)^2}
\end{equation}

and

\begin{equation}
\label{eq:dedt}
    \left.\frac{\mathrm{d}e_\mathrm{inner}}{\mathrm{d}t}\right|_\mathrm{GW} = -\frac{304}{15} \frac{\mathcal{G}^3}{c^5} 
    \frac{M_1 M_2 (M_1 + M_2)}{a_\mathrm{inner}^4}\frac{1 + \frac{121}{304}e_\mathrm{inner}^2}{\left(1 - e_\mathrm{inner}^2\right)^{5/2}}.
\end{equation}

\noindent We implemented Eqs.~(\ref{eq:dLdt}),~(\ref{eq:dedt}) into the \J{} code by including their contribution to the evolution of the Runge--Lenz and reduced orbital angular momentum vectors \citep[we refer to][for broader details about secularization]{Attia2021}. The outer orbit is assumed to emit a negligible amount of GWs, due to its much larger separation. Consistently with any hierarchical triple problem, the perturbing orbit hence acts as an angular momentum reservoir.

\subsection{Parameter space and MCMC sampling}
\label{subsec:mcmc}

We explored a seven-dimensional (7D) parameter space defined by the following:

\begin{itemize}
    \item Masses of the inner binary BHs: $M_1, M_2 \in [5,\,100]\,M_\odot$ with $M_2 \leq M_1$ (as the inner masses are interchangeable);
    \item Mass of the outer BH: $M_3 \in [1,\,200]\,M_\odot$;
    \item Semimajor axes: $a_\mathrm{inner} \in [1,\,200]\,\mathrm{AU}$, $a_\mathrm{outer} \in [100,\,10,000]\,\mathrm{AU}$ with $a_\mathrm{outer} > 10\,a_\mathrm{inner}$ (hierarchical approximation);
    \item Outer orbit eccentricity: $e_\mathrm{outer} \in [0,\,0.9]$. We excluded the $]0.9,\,1]$ range because we found a large discrepancy with $N$-body simulations for the highest eccentricities (Sect.~\ref{subsec:nbody});
    \item Mutual inclination: $i_\mathrm{mut} \in [40^\circ,\,80^\circ]$. We restricted our study to prograde orbits based on the well-established symmetry of ZLK dynamics about $90^\circ$, which holds when the angular momentum ratio of the inner and outer orbits is low \citep{Anderson2017}. This symmetry remains valid for ratios up to $\sim$\,0.1 \citep{Mangipudi2022}, and our hierarchical condition ensures that most of our parameter space falls well below this threshold, except near the boundaries where the inner and outer angular momenta can become commensurate. By excluding the region $i_\mathrm{mut} \in ]80^\circ,\,90^\circ]$, where we observed significant discrepancies with $N$-body simulations (Sect.~\ref{subsec:nbody}), we created two disconnected parameter space regions---the prograde regime we explore here and a roughly equivalent retrograde regime ($i_\mathrm{mut} \in [100^\circ,\,140^\circ]$) whose dynamics can be inferred through the symmetry relationship.
\end{itemize}

\noindent We did not vary the inner eccentricity $e_\mathrm{inner}$ because we know that if the ZLK mechanism is active, it will oscillate through its full range regardless of the initial value.

To efficiently sample the boundary between merging and nonmerging regions, we employed an MCMC approach. First, we simulated a grid of 300,000 systems spanning the parameter space. For each system, we computed whether it merged within 14\,Gyr or not, assigning a value of 1 (merger) or 0 (no merger). We then defined a scoring function for any point $\boldsymbol{x}$ in the parameter space based on its $k$ nearest neighbors in this grid:

\begin{equation}
\label{eq:f}
    f(\boldsymbol{x}) = \frac{1}{N}
    \sum_{\boldsymbol{x}_k\in\mathcal{N}_k(\boldsymbol{x})}{\frac{M(\boldsymbol{x}_k)}{d(\boldsymbol{x},\boldsymbol{x}_k)}}.
\end{equation}

\noindent Here $\mathcal{N}_k(\boldsymbol{x})$ is the set of $k = 100$ nearest neighbors to $\boldsymbol{x}$, $M(\boldsymbol{x})$ is 1 if system $\boldsymbol{x}$ merges and 0 otherwise, $d(\boldsymbol{x},\boldsymbol{y})$ is the standardized Euclidean distance between $\boldsymbol{x} = \left(x_1,x_2,...,x_7\right)$ and $\boldsymbol{y} = \left(y_1,y_2,...,y_7\right)$ such that 

\begin{equation}
    d(\boldsymbol{x},\boldsymbol{y}) = \sqrt{\sum_{i = 1}^7 \frac{\left(x_i - y_i\right)^2}{V(i)}},
\end{equation}

\noindent with $V(i)$ the variance of parameter $i$ over all computed points (to avoid the distance measure being dominated by the largest scales). Finally, $N$ is a normalization term:

\begin{equation}
    N = \sum_{\boldsymbol{x}_k\in\mathcal{N}_k(\boldsymbol{x})}{\frac{1}{d(\boldsymbol{x},\boldsymbol{x}_k)}}.
\end{equation}

\noindent This weighted average $f(\boldsymbol{x})$ (Eq.~(\ref{eq:f})) ranges from 0 to 1, representing the likelihood of merger for a system with parameters $\boldsymbol{x}$. We are most interested in regions where $f(\boldsymbol{x}) \simeq 0.5$, which corresponds to the boundary between merging and nonmerging configurations.

We convert $f(\boldsymbol{x})$ to a log-likelihood function $g(\boldsymbol{x})$ for the MCMC,

\begin{equation}
\label{eq:g}
    g(\boldsymbol{x}) = \log{\left[1 - \left(\frac{\lvert f(\boldsymbol{x})-0.5 \rvert}{0.5}\right)^2 \right]},
\end{equation}

\noindent which gives the highest scores to systems with $f(\boldsymbol{x}) \simeq 0.5$ and strongly penalizes systems with $f(\boldsymbol{x})$ near 0 or 1. We note that any strictly positive exponent inside the $\log$ function would make Eq.~(\ref{eq:g}) a valid log-likelihood function. We chose the value of 2 because it is the lowest exponent for which $g$ is $C^\infty$.

We ran the MCMC with 8,000 parallel workers for approximately 3,000 iterations each. Importantly, our approach deviates from the classical MCMC approach in a substantial way: as new samples are simulated during the MCMC process, they are added to the grid of precomputed systems used for scoring. This means the log-likelihood function $g(\boldsymbol{x})$ evolves throughout the process (since $\mathcal{N}_k(\boldsymbol{x})$ is refined, Eq.~(\ref{eq:f})) as more data become available, making our method self-improving. When a proposed sample is rejected, we still simulate its evolution and add the result to our database, thereby continuously enhancing the accuracy of our nearest-neighbor estimator. This adaptive approach is more effective than standard MCMC, as the $k$ nearest neighbors of a new sample become progressively closer and more representative as the grid density increases. In total, we simulated nearly 15 million different initial configurations; when duplicates are removed, we have 14,456,391 unique systems across all MCMC chains, including burn-in samples. After discarding the first 50\% of each chain as burn-in (to remove samples from the initial random exploration phase), we retained 14,006,724 samples in our final dataset, preserving duplicates. It is important to retain duplicates as they represent high-scoring regions that were sampled multiple times, and their frequency provides statistical weight in our analysis.

\subsection{Neural network for outcome prediction}
\label{subsec:nn}

To enable the rapid prediction of merger outcomes without full dynamical simulations, we trained a readily usable open-source neural network\footnote{\url{https://github.com/maraattia/TripleBHMergerPredictor}} on the results of our MCMC exploration. After removing duplicates, we had 14,456,391 unique systems with known outcomes, which we split into training (60\%), validation (20\%), and test (20\%) sets.

Our neural network architecture, typical of binary classification schemes, consists of
\begin{itemize}
    \item an input layer with seven neurons (one for each parameter);
    \item three hidden layers with 128 neurons each using ReLU activation functions;
    \item an output layer with a single neuron using a sigmoid activation function.
\end{itemize}

\noindent The input parameters were normalized to the range $[0,\,1]$. The network was trained for 15 epochs, at which point the validation accuracy no longer improved.

Although the model is trained on binary data, since the real outcome $y = M(\boldsymbol{x})$ is either 0 or 1, its predictions $\hat{y}$ are values between 0 and 1, which can be interpreted as the probability of merger. For binary classification, we use a threshold of 0.5, with outputs $\geq 0.5$ classified as mergers. We define in this framework a confidence measure

\begin{equation}
\label{eq:c}
    c(\hat{y}) = 2\lvert \hat{y} - 0.5 \rvert,
\end{equation}

\noindent which ranges from 0 (the model is completely uncertain of its prediction) to 1 (completely certain).

\section{Results}
\label{sec:results}

\subsection{Comparison with $N$-body simulations}
\label{subsec:nbody}

\begin{figure*}
    \centering
    \includegraphics[width=\linewidth]{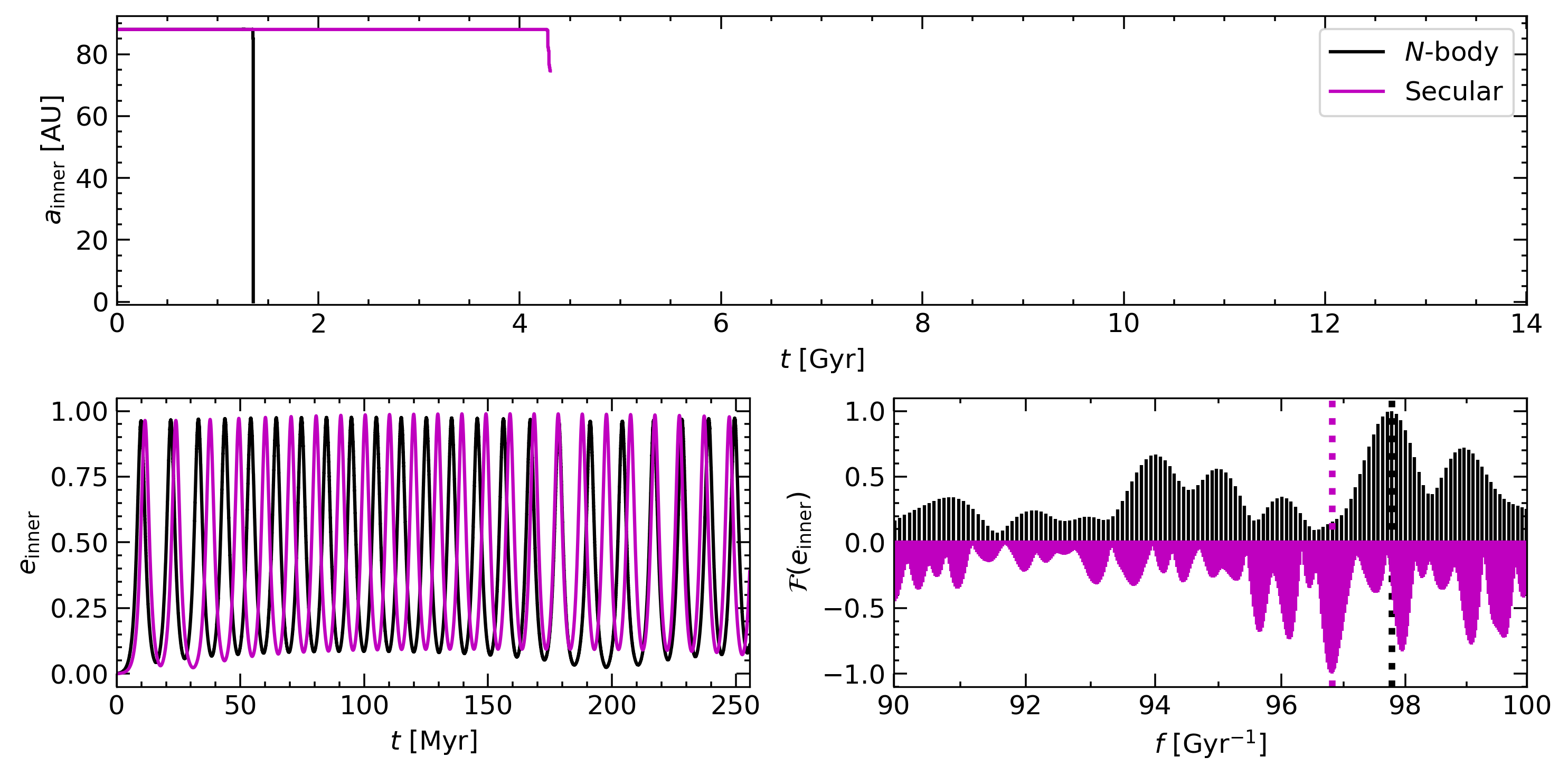}
    \caption{Comparative evolution of inner semimajor axis (top) and inner eccentricity (bottom left), with Fourier transform of inner eccentricity (bottom right) including position of peak frequencies (dotted lines), between $N$-body (black) and secular (purple) codes for two similar initial configurations: $M_1 = 68.7\,M_\odot$, $M_2 = 17.0\,M_\odot$, $M_3 = 92.0\,M_\odot$, $a_\mathrm{inner} = 88.0\,\mathrm{AU}$, $a_\mathrm{outer} = 6.82\,\mathrm{kAU}$, $e_\mathrm{outer} = 0.670$, $i_\mathrm{mut} = 78.3^\circ$.}
    \label{fig:sec_vs_nbody}
\end{figure*}

To validate our secular approach, we performed an extensive comparative analysis between the modified \J{} code and the \texttt{TSUNAMI} direct $N$-body integrator \citep{Trani2023}. To this end, our validation set comprised 1,000 systems, providing a statistically robust assessment of the secular approximation's reliability across different parameter regimes. We divided our validation sample into four categories:

\begin{itemize}
   \item Predictable outcomes (300 systems). We selected 150 systems expected to merge and 150 expected not to merge based on their location in well-characterized regions of parameter space. The secular and $N$-body codes agreed perfectly on the outcome (merger or no merger) for all 300 systems, confirming that our approach reliably captures the dynamics in regions far from the merger boundary.
   \item Uncertain outcomes (500 systems). For systems sampled from regions where the posterior merger probability was uncertain ($0.45 \leq f(\boldsymbol{x}) \leq 0.55$), we found good agreement between the two approaches. Of the 500 systems tested, 18 were dynamically unstable (the triple disintegrated during evolution). Among the remaining 482 stable systems, the secular code correctly predicted 382 outcomes (177 true positives and 205 true negatives), yielding a 79\% accuracy rate. The disagreements consisted of 69 false positives and 31 false negatives, with no strong bias toward either type of error.
   \item High mutual inclination regime (100 systems). Systems with $80^\circ < i_\mathrm{mut} < 90^\circ$ showed poor agreement between the two methods. Of 100 randomly sampled systems in this range, 12 were dynamically unstable, and among the remaining 88, only 45 (51\%) showed consistent outcomes, which motivated our decision to exclude this parameter region from our main analysis.
   \item High outer eccentricity regime (100 systems). Systems with $e_\mathrm{outer} > 0.9$ similarly showed poor agreement. Of the 100 systems tested, 37 were unstable; among the 63 stable systems, only 17 (27\%) showed consistent outcomes. The breakdown of the secular approximation at very high eccentricities likely occurs because the outer companion's close pericenter passages violate the assumption of well-separated orbital timescales.
\end{itemize}

Excluding the two problematic parameter regions (high $i_\mathrm{mut}$ and high $e_\mathrm{outer}$), our overall validation accuracy reaches 87\% for stable systems. This high agreement rate validates our methodology for large-scale parameter space exploration, while clearly delineating the regimes where the secular approximation should not be trusted. The dynamical instabilities we observed highlight a limitation of our simple hierarchical criterion ($a_\mathrm{outer} > 10\,a_\mathrm{inner}$). While conservative, this criterion does not account for the destabilizing effects of mass ratios and eccentricities. More sophisticated stability criteria exist in the literature \citep{Mushkin2020,Hayashi2022,Hayashi2023}, including recent machine learning approaches \citep{Lalande2022,Vynatheya2022}, which could better predict unstable configurations. However, our testing indicates that instabilities are relatively rare ($\sim$\,2\% of systems) outside the high $i_\mathrm{mut}$ and $e_\mathrm{outer}$ regimes, primarily occurring when $M_3 \gg (M_1 + M_2)$ \citep{Mardling2001}. Given our focus on mapping the merger boundary rather than predicting individual system evolution, we deemed this level of contamination acceptable.

For systems predicted to merge by both codes, there were notable quantitative differences in the merger timescales. The secular code typically predicted merger times that differed from $N$-body results by a factor of $2 - 3$, occasionally up to an order of magnitude, with the secular prediction generally being longer. This discrepancy arises primarily from differences in the predicted ZLK oscillation periods, with the secular code typically estimating slightly longer periods. Figure~\ref{fig:sec_vs_nbody} illustrates the comparative evolution of two similar systems as predicted by the two codes. While both codes agree that these systems will merge, the secular code predicts a slightly longer ZLK period. This is visible in the Fourier transform of the eccentricity oscillations, where the peak frequency is slightly lower for the secular code. Despite these phase differences, the qualitative behavior is consistent between the two approaches.

\subsection{Characterizing the parameter space}
\label{subsec:param_space}

\begin{figure*}
    \centering
    \includegraphics[width=\linewidth]{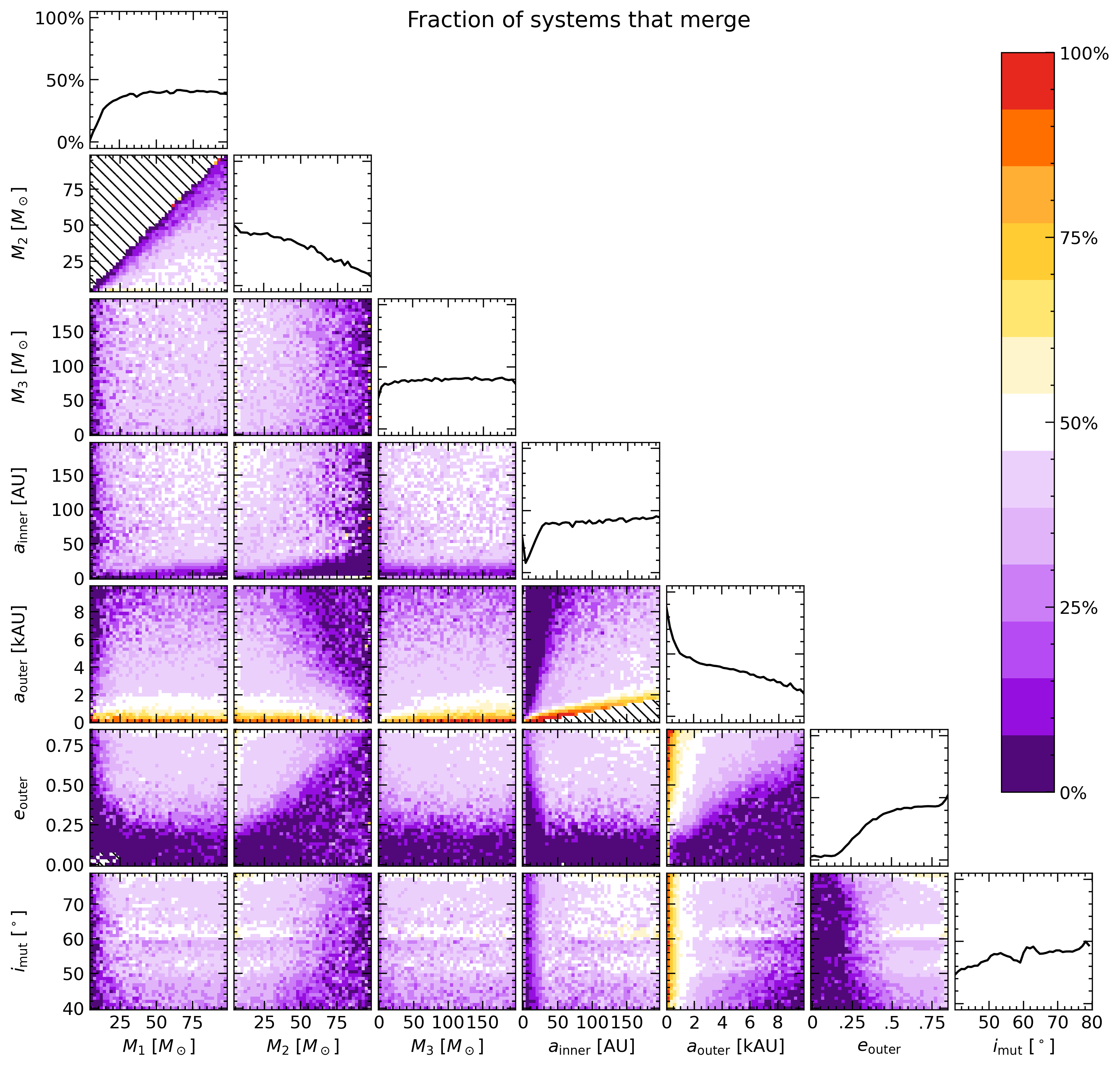}
    \caption{Fraction of systems that merge in each cell for each pair of parameters, or in each bin for each single parameter. The color ranges from purple (no systems merge) through white ($\sim$\,50\% merger rate) to yellow (all systems merge).}
    \label{fig:MCMC_distplot_mergerbool}
\end{figure*}

Our MCMC exploration efficiently identified the regions in parameter space where systems transition from certainly merging to certainly not merging. We monitored the Gelman--Rubin statistic for convergence and found $\hat{R} < 1.1$ for all parameters after burn-in. In regions with rapid transitions between outcomes (particularly near $i_\mathrm{mut} \simeq 60^\circ$, as developed below), we observed slower mixing, suggesting possible complex structure in the merger boundary. Our adaptive sampling approach partially mitigates this by continuously refining the likelihood surface. After analyzing approximately 15 million different initial configurations, we were able to characterize the merger probability distribution across the 7D parameter space and to identify key trends. Figure~\ref{fig:MCMC_distplot_mergerbool} shows the fraction of systems that merge as a function of individual parameters and parameter pairs. Merger-favorable regions are quite restricted, primarily concentrated around the specific parameter combinations that maximize the effectiveness of the ZLK mechanism or that enable direct merger via GW emission.

Examining the distribution of merger probability as a function of individual parameters reveals several important trends. For the inner binary masses ($M_1$ and $M_2$), we observe that the merger fraction is relatively flat with respect to $M_1$ for values above approximately $25\,M_\odot$. However, when $M_1 < 25\,M_\odot$, the merger fraction decreases, likely because both masses of the inner binary are low (due to our constraint that $M_2 \leq M_1$), resulting in weaker GW emission that cannot overcome the advantages of a stronger ZLK effect for lower masses. Notably, for a given total inner binary mass, a more asymmetric mass configuration ($M_1 \gg M_2$) shows a significantly higher merger fraction than a symmetric one. This can be traced to the octupolar contribution of the ZLK mechanism, which depends on the mass difference $M_1 - M_2$ \citep[e.g.,][]{Naoz2013}. A more asymmetric configuration can reach a higher maximum eccentricity, leading to a smaller pericenter distance and consequently a stronger GW emission. This finding challenges the intuition from isolated binary merger timescales, where equal masses typically lead to faster mergers at a given separation and total mass (Eq.~(\ref{eq:tgw})). Somewhat surprisingly, the tertiary mass ($M_3$) has minimal impact on merger probability across most of its range. Intuition might suggest that a more massive perturber would induce stronger ZLK oscillations, but our results indicate that this parameter is less critical than others in determining merger outcomes.

The semimajor axes ($a_\mathrm{inner}$ and $a_\mathrm{outer}$) show the most dramatic influences on merger probability. Systems with very small inner separations ($a_\mathrm{inner} < 10\,\mathrm{AU}$) can merge through GW emission alone, without significant assistance from ZLK oscillations. As $a_\mathrm{inner}$ increases beyond this threshold, systems increasingly require ZLK-induced high-eccentricity episodes to merge within the age of the Universe. The outer separation exhibits a strong negative correlation with merger probability: smaller $a_\mathrm{outer}$ values lead to much higher merger fractions, due to stronger perturbations on the inner binary. Systems with $a_\mathrm{outer} \gtrsim 10\,a_\mathrm{inner}$ (near the stability limit of hierarchical triples) almost universally merge. Furthermore, the outer eccentricity $e_\mathrm{outer}$ exhibits a strong positive correlation with merger probability. Higher eccentricities bring the tertiary closer to the inner binary at pericenter, enhancing the strength of ZLK oscillations and leading to higher inner binary eccentricities. This effect becomes particularly pronounced for $e_\mathrm{outer} > 0.6$. 

Similarly, the mutual inclination $i_\mathrm{mut}$ shows a generally positive correlation with merger probability, consistent with theoretical expectations that higher inclinations produce stronger ZLK effects \citep[e.g.,][]{Katz2011}. Interestingly, we observe a nonmonotonic feature with peaks in merger probability centered around $i_\mathrm{mut} = 54^\circ$ and $64^\circ$, separated by a trough at $i_\mathrm{mut} = 60^\circ$. This feature is statistically robust across different values of other parameters, but a clear physical explanation for this specific pattern remains elusive. It may in fact reflect the complex, potentially fractal nature of the boundary between merging and nonmerging configurations in the three-body problem \citep[in which regular and chaotic regions coexist at all scales, as shown by][]{Trani2024}. Additionally, as discussed by \citet{Hamers2021}, departing from the test-particle limit introduces asymmetries in eccentricity excitation, which can manifest as such irregular features.

\begin{figure*}
    \centering
    \includegraphics[width=\linewidth]{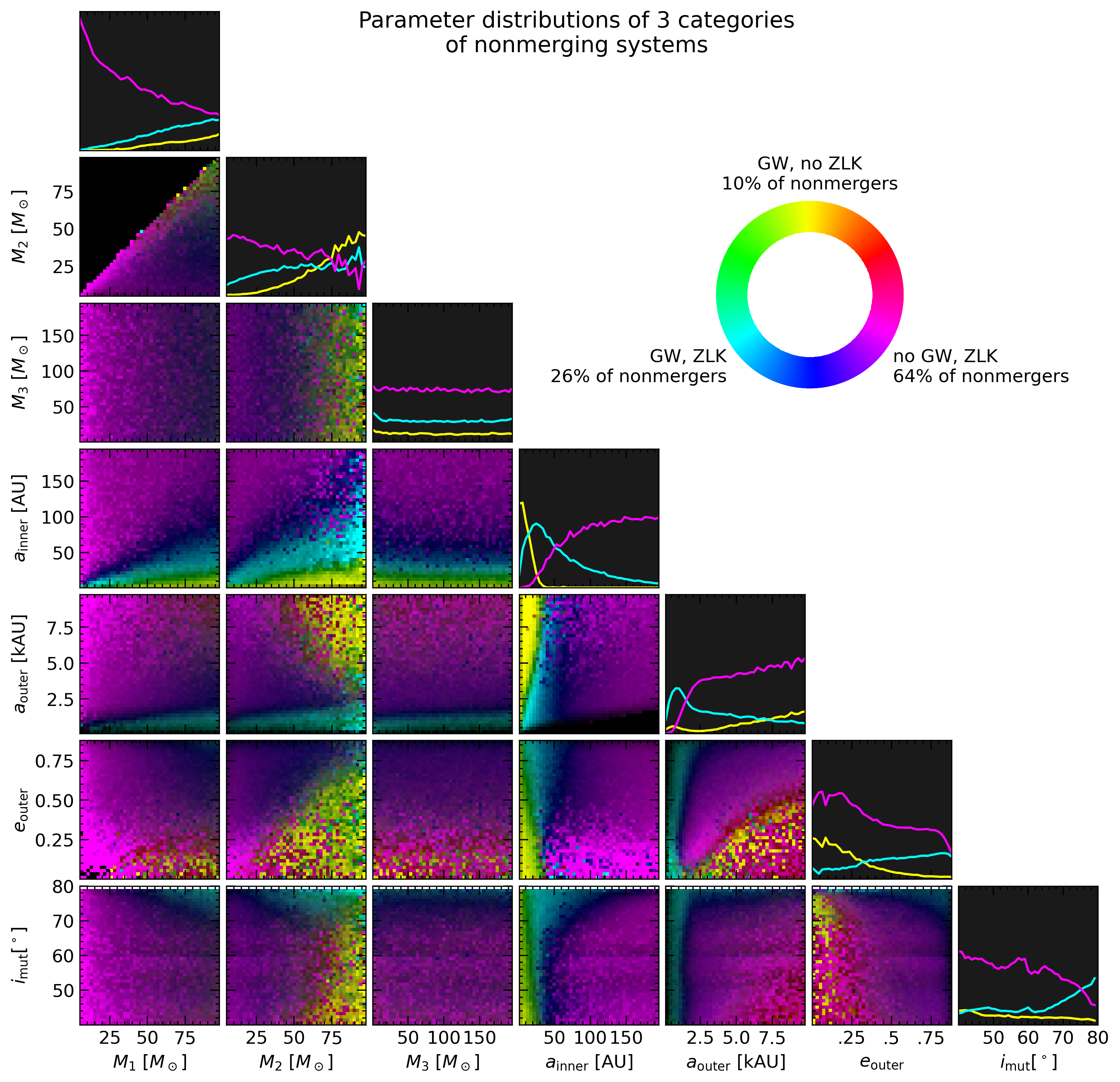}
    \caption{Distribution of three categories of nonmerging systems in the parameter space. Each category is represented by a base color: cyan (GW, ZLK), magenta (no GW, ZLK), and yellow (GW, no ZLK). For each pair of parameters, cells are color-coded according to the combination of the three base colors resulting from the fraction of systems corresponding to the three categories.}
    \label{fig:MCMC_cmy_distplot}
\end{figure*}

To further characterize the parameter space, we classified nonmerging systems based on their final state after evolving for 14\,Gyr. Specifically, we examined whether the inner binary's separation decreased (indicating GW emission) and whether the inner eccentricity changed from its initial value (indicating ZLK oscillations). This classification resulted in four distinct categories of nonmerging systems:

\begin{itemize}
    \item GW, ZLK (26\% of nonmerging systems): Systems that emit GWs and experience ZLK oscillations, but do not merge within 14\,Gyr. These systems are typically found in regions with moderate inner separations ($10 - 50\,\mathrm{AU}$) and outer separations smaller  than about 2,000\,AU.
    
    \item GW, no ZLK (10\%): Systems that emit GWs without significant ZLK oscillations. These are predominantly systems with very small inner separations ($< 10\,\mathrm{AU}$), where the binary is already close enough for GW emission to be significant, but where the ZLK mechanism is suppressed due to relativistic precession.
    
    \item No GW, ZLK (64\%): Systems that experience ZLK oscillations, but do not emit significant GWs. As expected, these systems have very low inner masses, or alternatively large inner separations where even the maximum eccentricity reached during ZLK cycles does not bring the pericenter close enough for substantial GW emission.
    
    \item No GW, no ZLK (<\,1\%): Systems that neither experience ZLK oscillations nor emit significant GWs. These rare cases typically occur when the mutual inclination is near the lower threshold for ZLK oscillations and other parameters are unfavorable.
\end{itemize}

Figure~\ref{fig:MCMC_cmy_distplot} visualizes the distribution of these categories across the parameter space, revealing clear regional trends. For each parameter, distinct patterns emerge in how they influence the distribution of nonmerging system types. The inner separation $a_\mathrm{inner}$ is particularly decisive. The ``GW, no ZLK'' systems dominate at very small separations ($a_\mathrm{inner} < 10\,\mathrm{AU}$), especially for far-orbiting perturbers (triggering minimal ZLK resonances) and for systems with more massive inner binaries, where GW emission is strong enough to drive orbital decay without requiring ZLK-induced eccentricity. As $a_\mathrm{inner}$ increases, we observe a transition to ``GW, ZLK'' systems, which peak at moderate separations ($10 - 50\,\mathrm{AU}$) where relativistic precession is weaker and allows ZLK oscillations to operate effectively, while still permitting significant GW radiation. At larger separations ($> 50\,\mathrm{AU}$), ``no GW, ZLK'' systems become prevalent, as the separations are too large for efficient GW emission regardless of ZLK-induced eccentricity. The outer separation $a_\mathrm{outer}$ also exhibits a clear gradient effect. When it is close to the stability limit ($a_\mathrm{outer} \gtrsim 10\,a_\mathrm{inner}$), systems predominantly merge (Fig.~\ref{fig:MCMC_distplot_mergerbool}). As $a_\mathrm{outer}$ increases, we first encounter a transition zone of mostly ``GW, ZLK'' systems, followed by a region with ``No GW, ZLK'' systems catching up at larger separations where the perturbing influence diminishes, reducing the maximum eccentricity achieved during ZLK cycles. 

The mutual inclination $i_\mathrm{mut}$ strongly influences the system category, with a noticeable pattern: ``no GW, ZLK'' systems prevail for the lowest inclinations, except when conditions are too unfavorable for ZLK (typically for circular perturbers and short inner orbits), and gradually fade out as $i_\mathrm{mut}$ is increased. This is related to the fact that the highest ZLK-induced eccentricities, required for efficient GW emission, can only be achieved with orbits that are tilted enough. Around $i_\mathrm{mut} \gtrsim 40^\circ$, ``GW, no ZLK'' systems are more common than elsewhere, as these inclinations are near the threshold required for ZLK oscillations. As inclination increases toward $90^\circ$, ``GW, ZLK'' systems become increasingly dominant, reaching over 50\% of nonmerging systems at the highest inclinations. Again, this reflects the greater efficiency of ZLK oscillations at architectures closer to perpendicular orbital configurations. For $e_\mathrm{outer}$, higher eccentricities show fewer ``No GW, ZLK'' and more ``GW, ZLK'' systems because the closer pericenter passages of the tertiary body enhance the GW radiation.

The masses of the three BHs undergo more subtle changes. For the inner binary masses $M_1$ and $M_2$, systems with higher masses show increased fractions of ``GW, no ZLK'' and ``GW, ZLK'' categories, as expected from the stronger GW emission from more massive binaries. However, the mass asymmetry in the inner binary is also important; when $M_1 \gg M_2$, the octupolar contributions to ZLK oscillations are enhanced, leading to higher fractions of systems experiencing ZLK oscillations. The mass imbalance effect is clear for the lowest mutual inclinations: such regions are usually populated by ``no GW, ZLK'' systems, but they transition to ``GW, no ZLK'' for the highest values of $M_2$. The latter correspond to zones where the octupolar effect is the most modest (due to our constraint that $M_2 \leq M_1$), hence suppressing ZLK cycles. The tertiary mass $M_3$ shows a relatively uniform distribution of categories across its range, confirming its secondary importance with respect  to geometric parameters. 

These detailed characterizations of the parameter space provide valuable insights into which initial configurations are most likely to produce BH mergers via the hierarchical triple channel. The clear regional patterns we identify can inform models of GW progenitor formation and help interpret the growing catalog of GW detections. We note that while we present here the results found using the natural mass and separation parameters, we verified that alternative parameterizations (e.g., $M_\mathrm{inner} = M_1 + M_2$, $q_{21} = M_2/M_1$, $q_3 = M_3/M_\mathrm{inner}$, $\alpha = a_\mathrm{inner}/a_\mathrm{outer}$) yield consistent conclusions about the key trends governing merger probability.

\subsection{Neural network performance}
\label{subsec:nn_results}

\begin{figure}
    \centering
    \includegraphics[width=\linewidth]{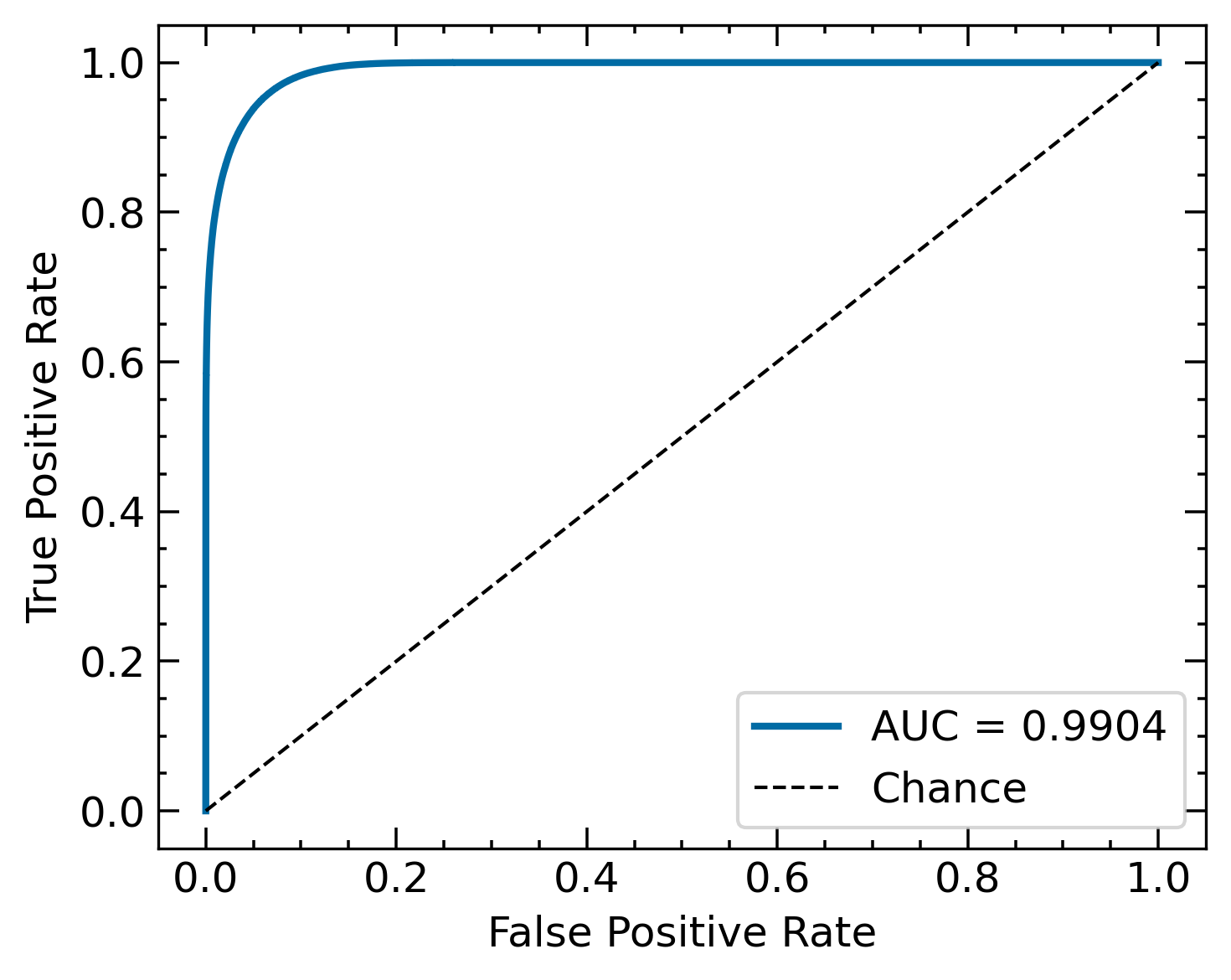}
    \caption{ROC curve of the neural network. The AUC is shown, as is the diagonal (pure chance).}
    \label{fig:roc_curve}
\end{figure}

The neural network we developed achieves an overall accuracy of 94.7\% on the test set of 2,632,159 samples (20\% of the full dataset), demonstrating its ability to reliably predict merger outcomes without performing full dynamical simulations. This high accuracy is particularly remarkable given the complex, nonlinear nature of the ZLK mechanism and its interaction with GW emission. Table \ref{tab:confusion_matrix} presents the confusion matrix of our model's predictions. The test set contains a somewhat unbalanced distribution, with approximately 70\% nonmerging systems, reflecting the natural distribution found while exploring the parameter space, including the rejected samples in the MCMC. Despite this imbalance, the model performs well on both classes, showing no critical differences between false positives (50,828) and false negatives (89,798).

\begin{table}
    \centering
    \begin{tabular}{ccccc}
    \multicolumn{5}{c}{\textbf{Actual Outcome}}\\
    & & No Merger & Merger & \textbf{Total} \\ \cline{3-5}
    \multirow{3}{*}{\textbf{Predicted}} & No Merger & \multicolumn{1}{|c|}{1,772,700} & \multicolumn{1}{|c|}{89,798} & \multicolumn{1}{c|}{1,862,498} \\ \cline{3-5}
    & Merger & \multicolumn{1}{|c|}{50,828} & \multicolumn{1}{|c|}{718,833} & \multicolumn{1}{c|}{769,661} \\ \cline{3-5}
    & \textbf{Total} & \multicolumn{1}{|c|}{1,823,528} & \multicolumn{1}{|c|}{808,631} & \multicolumn{1}{c|}{2,632,159} \\
    \cline{3-5}
    \end{tabular}
    \caption{Confusion matrix of the neural network on the test set.}
    \label{tab:confusion_matrix}
\end{table}

The model's performance, illustrated by the receiver operating characteristic (ROC) curve in Fig.~\ref{fig:roc_curve}, can be characterized by several standard metrics:
\begin{itemize}
    \item Positive predictive value (precision): 93.4\%---when the model predicts a merger, it is correct 93.4\% of the time;
    \item Negative predictive value: 95.2\%---when the model predicts no merger, it is correct 95.2\% of the time;
    \item True positive rate (recall): 88.9\%---the model correctly identifies 88.9\% of all merging systems;
    \item True negative rate (specificity): 97.2\%---the model correctly identifies 97.2\% of all nonmerging systems;
    \item F1 score: 91.1\%---the harmonic mean of precision and recall, providing a balanced measure of the model's performance;
    \item Area under the ROC curve (AUC): 99.0\%---this metric reflects the model's ability to distinguish between merging and nonmerging classes across all classification thresholds.
\end{itemize}

\begin{figure}
    \centering
    \includegraphics[width=\linewidth]{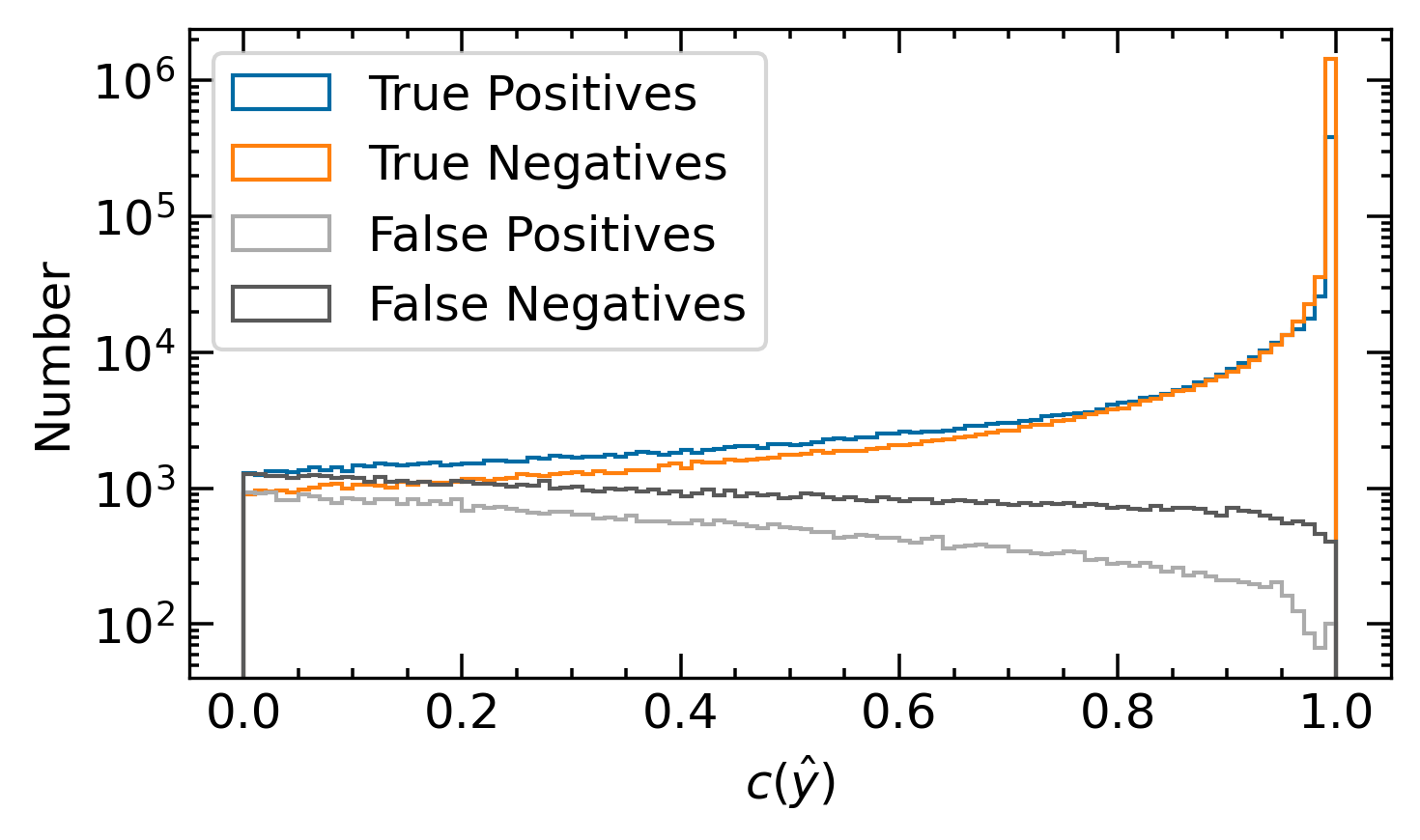}
    \caption{Histograms of true positives, true negatives, false positives, and false negatives as functions of prediction confidence $c(\hat{y})$.}
    \label{fig:certainty_histogram}
\end{figure}

A particularly valuable aspect of our neural network implementation is the confidence measure (Eq.~(\ref{eq:c})), which provides a robust assessment of prediction reliability. This measure allowed us to analyze the relationship between the model's confidence and its accuracy, revealing a strong positive correlation. Figure~\ref{fig:certainty_histogram} illustrates this relationship by showing the distribution of true and false predictions across different confidence levels. Analysis of the confidence distribution reveals that 69.6\% of all test samples have a confidence measure above 0.99, and 79.5\% have confidence above 0.9. More importantly, the accuracy is strongly correlated with confidence: for predictions with $c(\hat{y}) > 0.9$, the accuracy increases dramatically to 99.7\%. This includes 69.3\% of true positive and 89.4\% of true negative predictions having confidence above 0.9, while only 3.0\% of false positives and 6.5\% of false negatives have such high confidence. For these high-confidence systems, the model achieves a positive predictive value of 99.7\% and a negative predictive value of 99.6\%. In practical terms, this means that for approximately 80\% of all systems, we can predict the merger outcome with near certainty. For the remaining 20\% of systems with lower confidence predictions, the model still achieves approximately 75\% accuracy.

The singular performance of our neural network, particularly its ability to provide reliable confidence measures, makes it an ideal tool for rapid population synthesis studies. Researchers can use this model to quickly classify millions of systems, focusing detailed simulations only on the subset of cases where the neural network indicates uncertainty. The prediction phase is remarkably efficient, requiring only milliseconds per system or a few seconds for millions of systems on standard computational hardware, representing an improvement of many orders of magnitude compared to full simulations.

\section{Discussion and conclusion}
\label{sec:conclusion}

Our study has several caveats that should be acknowledged. While computationally efficient, our secular approach has inherent limitations. First, we do not include semisecular corrections that become important in the most compact hierarchical configurations \citep[as reviewed by][]{Tremaine2023}. These corrections, particularly the single-averaged terms that account for variations on the inner orbital timescale, can significantly enhance eccentricity excitation and potentially reveal additional merging systems \citep{Mangipudi2022}. Second, our validation against $N$-body simulations revealed poor accuracy of the secular formalism for near-perpendicular mutual inclinations on the one hand, and for near-unity outer eccentricities where the fundamental assumption of well-separated orbital timescales is violated on the other hand. The discrepancies in these regimes may partly stem from the missing semisecular corrections becoming crucial when orbital timescales approach commensurability. However, these same regions also exhibit high rates of dynamical instability (12\% for $i_\mathrm{mut} \in [80^\circ,\,90^\circ]$ and 37\% for $e_\mathrm{outer} > 0.9$), making their exclusion from our parameter space prudent regardless of the theoretical framework employed. Third, our simple hierarchical stability criterion permits some ($\sim$\,2\%) dynamically unstable configurations in the remaining parameter space. While more sophisticated stability criteria incorporating masses and eccentricities are available, we opted for computational efficiency given that our primary goal is mapping broad merger probability trends rather than predicting individual system evolution. 

Despite these limitations, our 87\% agreement rate with $N$-body simulations (excluding the problematic high $i_\mathrm{mut}$/$e_\mathrm{outer}$ regimes) confirms that the secular approximation captures the essential dynamics for the vast majority of our parameter space. Nevertheless, it should be noted that we left several parameters unexplored. We actually did not vary the arguments of pericenter or longitudes of ascending node. BH spins were also not included in our dynamical model, though spin--orbit and spin--spin couplings could influence evolution in some systems \citep{Kidder1993,Hartl2005,Bern2021}. In addition, we started our simulations after BH formation, ignoring the effects of stellar evolution and the ZLK mechanism during the progenitor phase. Finally, our simulations did not follow the final inspiral phase in detail, thus preventing accurate gravitational waveform predictions.

Still, we   conducted in this work the most comprehensive exploration to date of the hierarchical triple BH parameter space, analyzing the conditions under which such systems can produce inner binary mergers within the age of the Universe through an intensive study scrutinizing around 15 million configurations. Our investigation reveals that the ZLK mechanism provides a viable and efficient pathway for overcoming the initial separation problem in compact object binaries, particularly under specific parameter combinations that maximize the mechanism's effectiveness. The general patterns we were able to pinpoint are that systems with asymmetric inner binary masses benefit from enhanced octupolar contributions to the ZLK mechanism, leading to stronger eccentricity oscillations and consequently more efficient GW emission. When combined with moderate inner separations (where relativistic precession does not overwhelm the ZLK effect), small outer separations (providing stronger perturbations), and high outer eccentricities (bringing the perturber closer at pericenter), these configurations consistently produce mergers within cosmological timescales.

The limit between merging and nonmerging configurations in our 7D parameter space exhibits complex structures. While we can identify broad statistical trends, the three-body problem's inherently chaotic nature means this boundary is not smooth, but likely exhibits fractal characteristics. As demonstrated by \citet{Trani2024}, regular and chaotic trajectories are interwoven at all scales in a multifractal pattern, with regular regions occupying a substantial portion of the phase space depending on the initial configuration. In the hierarchical limit with quadrupole approximation, the system can be integrable \citep[e.g.,][]{Kinoshita1999}, but relaxing these assumptions (as done in this work) introduces chaos that may manifest as fractal boundaries between different outcomes. 

Our methodological approach, coupling secular approximation with an adaptive MCMC technique, has globally proven effective for navigating this high-dimensional parameter space. By targeting the transition regions between certainly merging and certainly nonmerging configurations, we   mapped the probability landscape in fine detail. This mapping not only characterizes where mergers occur, but also identifies distinct categories of nonmerging systems, providing additional insights into the interplay between ZLK oscillations and GW emission. Nonetheless, the true boundary likely has a fractal dimension smaller than the embedding dimension of our parameter space. This means that arbitrarily fine sampling would continue to reveal new structure, with pockets of regular behavior embedded within chaotic regions and vice versa. Future work could investigate the fractal dimension of these boundaries and their impact on merger rate predictions.

Perhaps the most practical outcome of our work is the neural network model,\footnote{\url{https://github.com/maraattia/TripleBHMergerPredictor}} which provides very fast predictions of merger outcomes with good accuracy. This tool transforms what would otherwise be computationally prohibitive population synthesis studies into feasible investigations. With an area under the ROC curve of 99\%, 95\% overall accuracy, and nearly perfect predictions (99.7\%) for the majority of systems, researchers can now efficiently identify promising configurations for further detailed study without the computational burden of full dynamical simulations.

As GW detectors continue to increase in sensitivity and the catalog of detected signals grows, the patterns we have identified can help constrain the astrophysical origins of observed mergers. Our results suggest that hierarchical triple systems contribute a distinct population of merging BHs, potentially distinguishable by their mass ratios and orbital characteristics from those produced by other formation channels. Future work connecting our parameter space exploration to realistic initial conditions from stellar evolution models will further refine our understanding of the relative importance of this pathway in nature. Ultimately, this study advances our understanding of the complex gravitational dynamics that shape the Universe's most extreme environments, bringing us closer to deciphering the origin stories written in the GWs that ripple through spacetime.

\begin{acknowledgements}
We are grateful to our referee, Alessandro Alberto Trani, for improving the quality of our article through a thoughtful review. We offer our thanks to Georges Meynet for his support and for fruitful conversations. We made use of the Claude AI assistant (Anthropic, 2024) for code optimization. We thank the University of Geneva HPC team for computational resources.
\end{acknowledgements}

\bibliographystyle{aa} 
\bibliography{biblio} 

\end{document}